# Observation of Faraday rotation from a single confined spin

Mete Atatüre[1*†], Jan Dreiser[1*], Antonio Badolato[1], Atac Imamoglu[1†]

[1] Institute of Quantum Electronics, ETH Zurich, CH-8093 Zurich, Switzerland.

[*] These authors contributed equally to this work.

[†] To whom correspondence should be addressed. Email: atature@phys.ethz.ch and imamoglu@phys.ethz.ch

**Ability to read-out the state of a single confined spin lies at the heart of solid-state quantum information processing[1]. While all-optical spin measurements using Faraday rotation has been successfully implemented in ensembles of semiconductor spins[2-4], read-out of a single semiconductor spin has only been achieved using transport measurements based on spin-charge conversion[5,6]. Here, we demonstrate an all-optical dispersive measurement of the spin-state of a single electron trapped in a semiconductor quantum dot. We obtain information on the spin state through conditional Faraday rotation of a spectrally detuned optical field, induced by the polarization- and spin-selective trion (charged quantum dot) transitions. To assess the sensitivity of the technique, we use an independent resonant laser for spin-state preparation[7]. An all-optical dispersive measurement on single spins has the important advantage of channeling the measurement back-action onto a conjugate observable, thereby allowing for repetitive or continuous quantum nondemolition (QND)[8] read-out of the spin-state. We infer from our results that there are of order**

**unity back-action induced spin-flip Raman scattering events within our measurement timescale. Therefore, straightforward improvements such as the use of a solid-immersion lens[9,10] and higher efficiency detectors would allow for back-action evading spin measurements, without the need for a cavity.**

Absorption and dispersion coexist in an optical field's response to a spectrally detuned optical transition. While these responses are of comparable strength for small detunings, dispersive response dominates over the absorptive part as the spectral detuning is increased. Measurement of the dispersive response could provide information about the ground (spin) state, if the transition at hand is spin-selective with definite optical selection rules, as is the case for a quantum dot (QD) confining a single excess electron.[7,11] In this Letter, we demonstrate a measurement of a QD spin by detecting this dispersive response in the form of Faraday rotation of a far-detuned linearly polarized optical field. Since the measurement field is detuned by as much as 340 times the transition linewidth from QD resonances, the dispersive response dominates over that of absorptive.

Theoretical proposals based on Faraday rotation from a QD embedded in a microcavity have suggested that QND measurement of a single spin could be implemented[12]. Remarkably, our observations suggest that the back-action evading spin measurement could be realized even in the absence of an optical cavity enhancing the Faraday rotation: while we estimate that the QD scatters a photon every 6 microseconds, the role of these photons is to leak information about the spin state into the radiation field reservoir without inducing a back-action on electron spin. The spin-flip Raman scattering events, which provide a back-action channel, occur once every 60 milliseconds. While it is impossible to avoid photon scattering within a measurement time yielding a Faraday signal-to-noise (SNR) ratio exceeding unity, spin-flip scattering can be negligible provided that the ratio of the peak absorption cross-section to the laser focal spot area exceeds the branching ratio to the spin-flip scattering channel ($< 10^{-3}$ in self-assembled QDs under 1-Tesla external magnetic field).

A single electron confined in a QD presents a four-level system in the trion representation as illustrated in Fig. 1A. The ground state is one excess electron in $|\downarrow\rangle$ ($|\uparrow\rangle$) state. The excited state $|\uparrow\downarrow, \blacktriangledown\rangle$ ($|\uparrow\downarrow, \blacktriangle\rangle$) corresponds to the QD with two electrons forming a singlet and a hole with angular momentum projection $J_z = -3/2$ ( $3/2$ ) along the growth direction. The trion transition, $|\uparrow\downarrow, \blacktriangledown\rangle \longrightarrow |\downarrow\rangle$ ($|\uparrow\downarrow, \blacktriangle\rangle \longrightarrow |\uparrow\rangle$), is allowed only for $[\sigma^{(-)}]$ ($[\sigma^{(+)}]$) circular polarization as determined by the optical selection rules[13]. If a linearly polarized optical field, e.g. $[\pi^{(+)}]$, feels this transition, the $[\sigma^{(-)}]$ ($[\sigma^{(+)}]$) component of the polarization acquires a phase shift rotating the optical field's polarization by an angle $\theta$ (-$\theta$) in the linear basis. Owing to the principle of Pauli Blockade[14], only one of these transitions is available at any given time and the laser polarization is rotated either in positive or negative angular direction conditional on the spin state of the confined electron. This conditionality links the electron spin to the light polarization directly even when excited-state population is negligible. In our results we exploit precisely this spin-state-dependence of the optical field's dispersive response.

Figure 1B is an illustration of our detection scheme. A polarized laser is focused onto and recollected from a gated heterostructure incorporating a sparse density of QDs[5]. The polarizing beam splitter distributes the transmitted optical field into two linear polarization components in the rectilinear basis of (X, Y) with respect to $[\pi^{(+)}]$ and directs each arm to a photodiode. Along with independent access to each detector's output, such a configuration allows us to measure their sum and difference corresponding to the purely absorptive and purely dispersive response, respectively, all in one run. The experimental data presented in the following figures is this difference measurement, unless stated otherwise.

Under an external magnetic field, the Zeeman splitting lifts the spectral degeneracy of the two trionic transitions (as is the case in Fig. 1A). Consequently the dispersive response, which is otherwise cancelled due to the degeneracy and fast spin-flips induced by the hyperfine interaction[15], is expected to appear along with the absorptive part in the vicinity of each transition. Figure 1C shows how the sum (black circles) and difference (red circles) signals behave as a function of laser polarization basis, when an external magnetic field of 1 Tesla is applied along the strong confinement axis of the quantum dot. The trionic transitions in this case are split by 26 GHz, and the optical field's response in the near vicinity of the [$\sigma^{(-)}$]-polarized transition linked to spin-down electronic state is displayed. When the laser is also circularly polarized (left plot), it acquires an overall phase that can not be detected leading to a purely absorptive signal (black circles). Same scan with a linearly polarized laser (right plot) displays the dispersive response (red circles) alongside the absorptive (black circles): since the acquired phase is now a relative one, it leads to a rotation of the linear polarization. At such small detuning with respect to Zeeman splitting, the laser experiences Faraday rotation primarily due to one Zeeman transition nearby.

We now consider probe-laser detunings that are larger than the Zeeman splitting, where the difference signal we observe is due to the competition between the two transitions. Figure 2A shows the difference signal for the probe laser detuning from 30 GHz to 45 GHz with respect to the [$\sigma^{(+)}$]-polarized trion transition obtained in a 60-sec measurement timescale per point. The black circles display the difference (offset) signal when the preparation laser is left detuned from either of the two Zeeman transitions: in this case, no state-preparation is implemented and the electron spin-state remains close to a completely mixed state[16]. As the gate voltage is increased, the two Zeeman-split optical transitions experience an equal strength of DC-Stark shift. Consequently, the detuning of the probe laser with respect to the two transitions is also decreased, creating an increasing offset signal in accordance with the incommensurate detunings and *partial* cancellation of the Faraday rotations. The red circles display the difference signal when the preparation laser is in the near-vicinity of the [$\sigma^{(-)}$]-polarized Zeeman transition. At 415 mV gate voltage the preparation laser hits the resonance and the electron is spin cooled to the spin-up state with near-unity fidelity due to state-mixing induced spin-flip Raman transitions[7]. Therefore, we no longer observe the difference of two dispersive responses, but rather the full signal due to one spin-state. Figure 2B is essentially the same measurement when the preparation laser is tuned to resonance with the [$\sigma^{(+)}$]-polarized Zeeman transition again at 415 mV gate voltage, preparing the electron in the spin-down state. The full dispersive signal from a single electron spin is recovered, but now with opposite sign indicating Faraday rotation in the opposite direction. The amplitude of this signal is less than that of Fig. 2A in accordance with the additional detuning of 26 GHz due to Zeeman splitting. Beyond 450 mV gate voltage, the quantum dot charging state switches from one excess electron to two electrons forming a singlet spin state. Since the trionic transitions are no longer present beyond this point, the optical field experiences no dispersive response.

Figure 2C is a Faraday rotation angle plot for the probe laser as a function of preparation laser detuning with respect to the Zeeman transitions. The red (blue) circles at preparation-laser resonance correspond to the probe laser's polarization rotation angle in response to spin-up (spin-down) prepared state at ~ 100 $\Gamma$ (~ 185 $\Gamma$). The red (blue) squares correspond to the Faraday rotation angle for a probe detuning of ~ 220 $\Gamma$ (~ 306 $\Gamma$). The gray circles and squares indicate Faraday rotation angle when the spin is not prepared in a particular state in either of the above-mentioned detunings. The white circles show that there is no Faraday rotation when the QD has a spin singlet of two electrons.

Figure 3A displays a full map of the dispersive signal at a 100-msec measurement timescale, plotted as a function of gate voltage and probe-laser detuning when the preparation laser is [$\sigma^{(+)}$]-polarized. The center frequency of the probe laser is detuned 92 GHz (~ 306 times the transition linewidth) for the top figure and 56 GHz (~ 185 times the transition linewidth) for the bottom figure. At a gate voltage of 415 mV the preparation laser becomes resonant and the electron is prepared in spin-down state. The dispersive signal is visible even at this measurement time scale. Above 450 mV there is no signal, since the charging state of the quantum dot is a two-electron singlet configuration and dispersive response disappears. Figure 3B is a similar measurement when the preparation laser is [$\sigma^{(-)}$]-polarized. Due to Zeeman splitting of the two transitions, the center frequency of the probe laser is detuned 66 GHz (~ 220 times the transition linewidth) for the top figure and 30 GHz (~ 100 times the transition linewidth) for the bottom figure. Once more at 415 mV the preparation laser becomes resonant and the electron is prepared in the spin-up state with near-unity fidelity. The signal from a spin in a mixed state remains identical, while the signal from the optically prepared spin switches sign. The middle plot serves to display line projections of the signal strength on resonance for both spin states. The decrease in signal level is in accordance with the expected inverse detuning ($\delta^{-1}$) dependence.

If, during a spin measurement, back-action occurs on a timescale faster than the natural spin-flip times, then the measured dynamics will be distorted by the back-action even when a short read-out timescale is obtained with respect to natural spin-flip dynamics, i.e. spin $T_1$ time. Therefore, while the measurement time can speed up with technical improvements, it is essential that the back-action remains negligible within the spin $T_1$ time. We infer that the back-action of our interaction mechanism within the relevant timescale of natural spin-flip events is indeed negligible. However, we do not claim a quantum nondemolition or back-action evading measurement here, since during the required measurement time to obtain a unity SNR level, between 1 to 10 back-action events occur (for a branching ratio of $10^{-4}$ and $10^{-3}$, respectively). This limitation arises from the fact that our optical system has a numerical aperture of 0.5 and a photo-detector efficiency of 10%. By using a combination of index-matched solid-immersion lenses and commercially available higher efficiency photo-detectors, it would be possible to effectively eliminate measurement back-action in the form of spin-flip Raman scattering from the probe laser. In fact, based on the anticipated phonon[17,18] and co-tunneling[19] limited spin-flip times, we could envision resolving spin quantum jumps. Embedding the QD in a microcavity is an alternative strategy that has been previously proposed[12]:

incorporating gated structures into photonic crystal nano-cavities[20] is demanding, but the existence of a far-detuned cavity mode can well be the way to obviate measurement back-action.

21. Supported by NCCR Quantum Photonics. The Authors acknowledge many useful discussions with G. Giedke, A. Högele, C. Galland, G. Fernandez, J. Taylor and G. Salis. M. A. and J. D. would also like to thank J. Cash for technical assistance.

**Figure Captions**

**Fig. 1. (A)** Four-level scheme illustrating the ground and excited states of a single-electron-charged quantum dot (QD). Each ground state is linked to an excited state through circular polarization due to the optical selection rules. An optical field in the vicinity of a QD transition will experience a polarization rotation due to Faraday Effect. **(B)** An illustration of the experimental apparatus. A laser beam is impingent on the sample with a single QD in the focused laser area. Upon transmission, the laser is distributed by a polarizing beam splitter to two detectors. **(C)** Differential transmission of a laser field through a single-electron-charged QD at 1-Tesla external magnetic field. The left (right) figure is obtained using circularly (linearly) polarized laser. The black circles represent the sum of two detector signals proportional to the absorptive response, while the red circles represent the difference of the two detector signals proportional to the dispersive response. The probe laser power is 20 nW corresponding to a Rabi frequency $\Omega_L \sim \Gamma$ on resonance, and the relative value of absorption is 0.15% for circular polarized excitation at this power level. The data is obtained in the cotunneling regime to avoid spin pumping.

**Fig. 2. (A)** Dispersive signal from a probe laser detuning from 30 GHz to 45 GHz with respect to the [$\sigma^{(+)}$]-polarized trion transition, with a laser power of 1 μW. The black circles correspond to the signal when the preparation laser is left detuned from either of the two Zeeman transitions, and has no effect on the electron spin. The red circles correspond to the signal when the preparation laser is in the vicinity of the red Zeeman transition. At 415 mV gate voltage the preparation laser hits resonance with the [$\sigma^{(-)}$]-polarized Zeeman transition and the electron is spin cooled to the spin-up state with near-

unity fidelity. **(B)** The preparation laser in the vicinity of the [$\sigma^{(+)}$]-polarized Zeeman transition. Again at 415 mV gate voltage, the electron is prepared in the spin-down state. The dispersive signal from a single electron spin is recovered, but with opposite sign indicating the change of direction of Faraday rotation. **(C)** Faraday rotation angle proportional to spin-state occupation for the probe laser at two detunings as a function of preparation laser detuning with respect to the Zeeman transitions. The red (blue) circles correspond to the probe laser's polarization rotation angle in response to spin-up (spin-down) prepared state at ~ 100 Γ (~ 185 Γ). The red (blue) squares correspond to the Faraday rotation angle for a probe detuning of ~ 220 Γ (~ 306 Γ). The gray circles (squares) indicate Faraday rotation angle when the spin is not prepared in a particular state. The white circles indicate the signal level when the QD has a spin singlet of two electrons.

**Fig. 3. (A)** A full map of the dispersive signal as a function of gate voltage and detuning of a [$\pi^{(+)}$]-polarized probe laser, for two different detuning ranges obtained with a measurement timescale of 100 msec. The preparation laser hits resonance at 415 mV as indicated by the blue dashed line, preparing the electron in the spin-down state. **(B)** Similar measurements when the electron is prepared in the spin-up state via optical pumping on the [$\sigma^{(-)}$]-polarized Zeeman transition. The dispersive signal strength at 415 mV is displayed in the middle plot as a cut along the blue (spin-down) and red (spin-up) dashed lines along with the theoretically expected behavior.

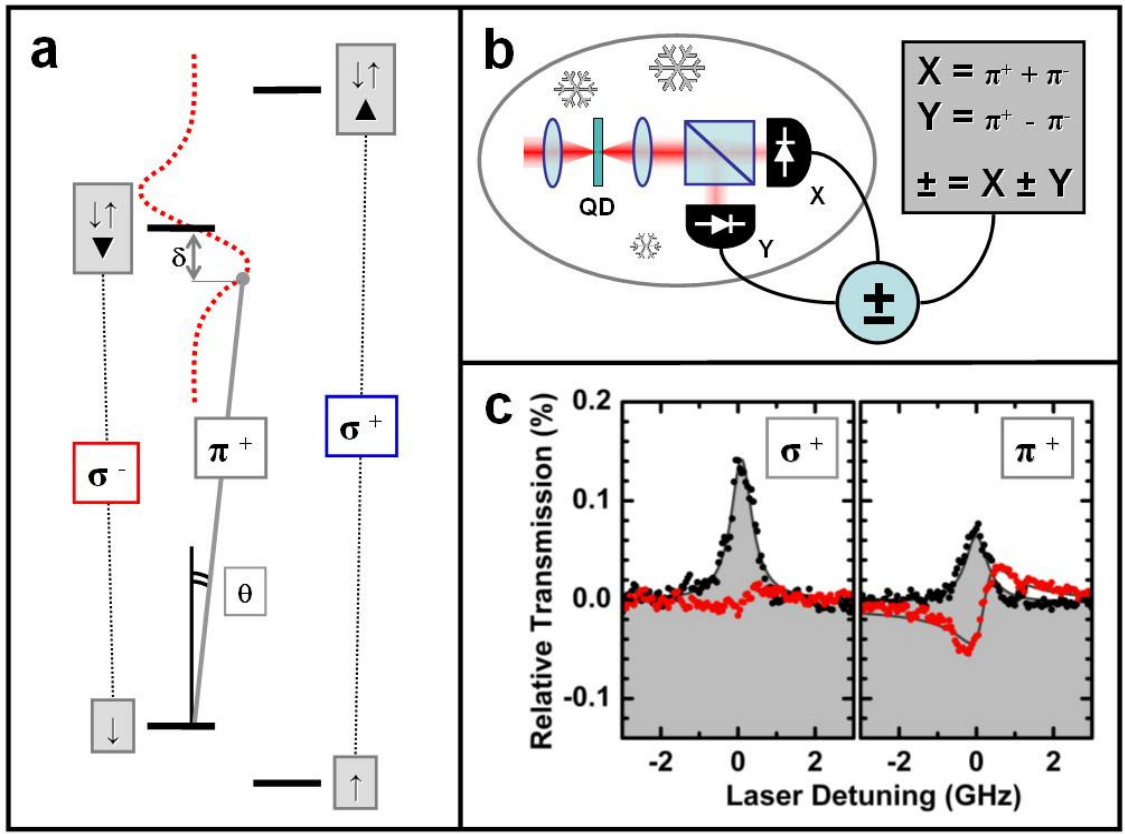

**Figure 1**

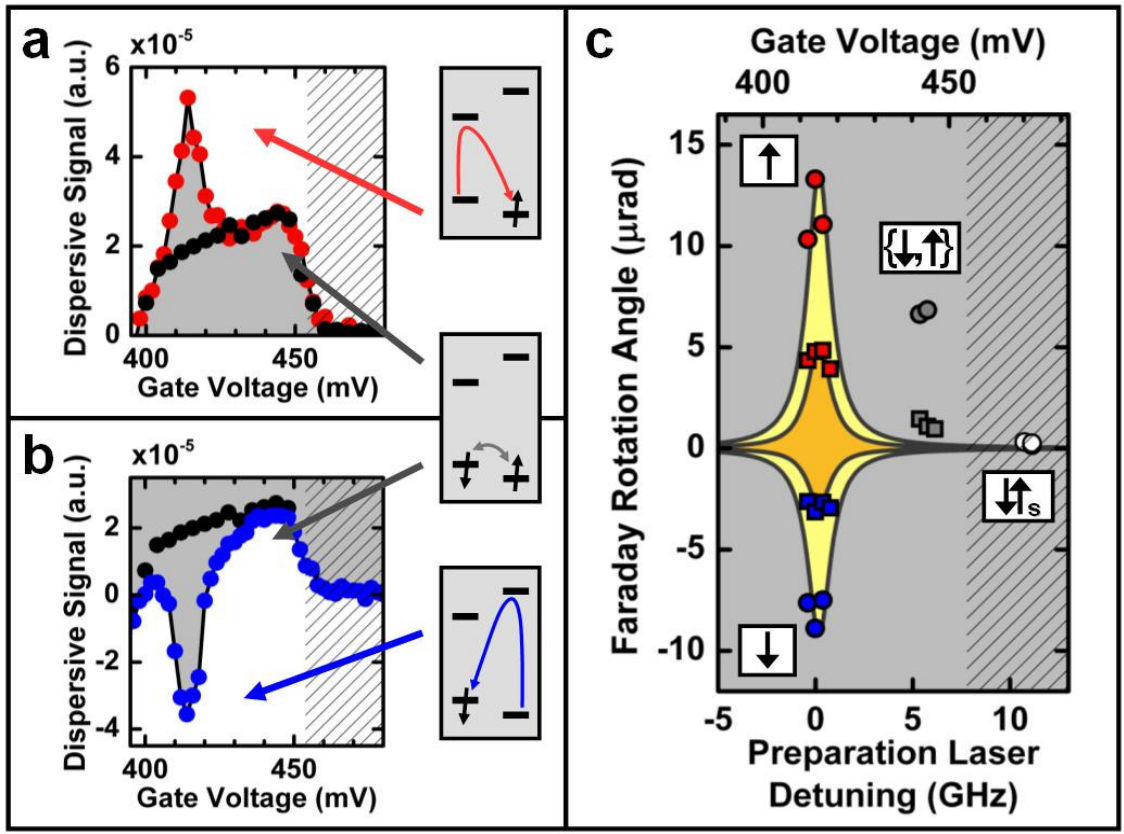

**Figure 2**

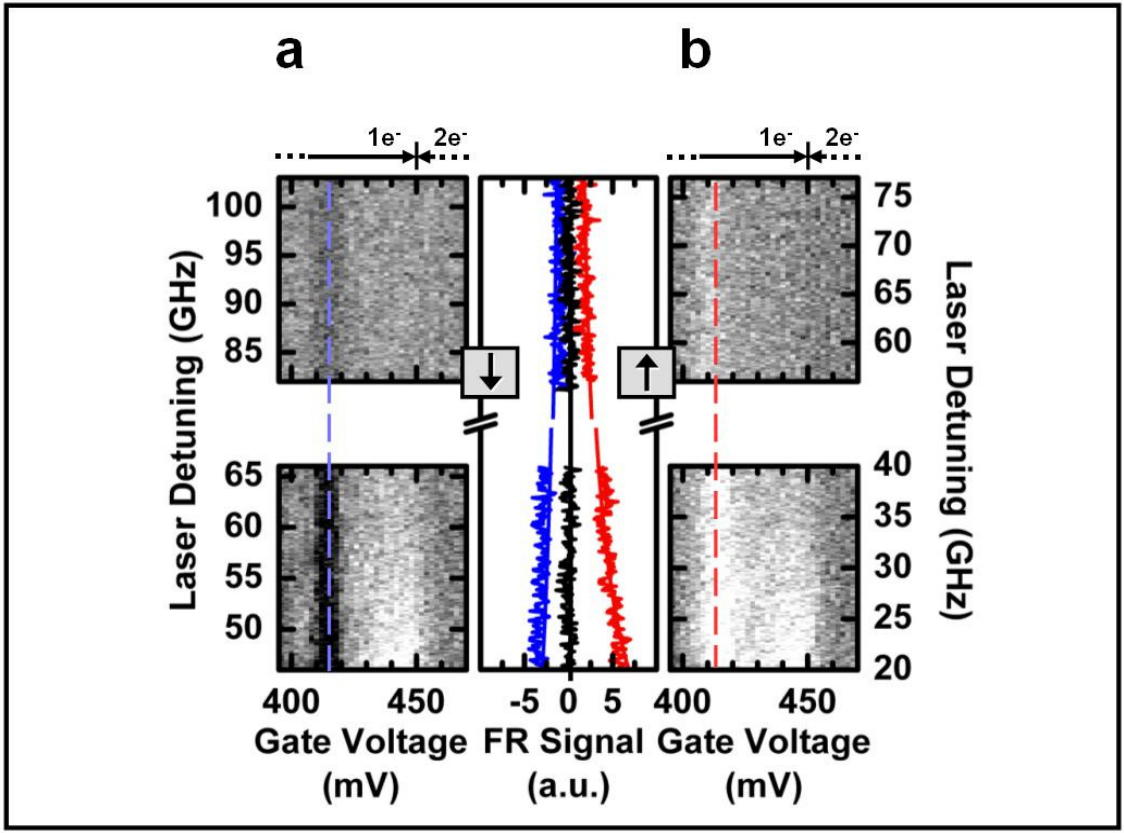

**Figure 3**